\documentclass[twocolumn,showpacs]{revtex4}
\usepackage{amssymb}
\usepackage{makeidx}
\usepackage{amsmath}
\usepackage{graphicx}
\usepackage{dcolumn}
\usepackage{bm}
\usepackage{epstopdf}

\begin{document}

\title{Enhanced harmonic generation and wave-mixing via two-color
multiphoton excitation of atoms/molecules}
\author{H. K. Avetissian}
\author{B. R. Avchyan}
\author{G. F. Mkrtchian}
\affiliation{Centre of Strong Field Physics, Yerevan State University, 1 A. Manukian,
Yerevan 0025, Armenia }
\date{\today }

\begin{abstract}
We consider harmonics generation and wave-mixing by two-color multiphoton
resonant excitation of three-level atoms/molecules in strong laser fields.
The coherent part of the spectra corresponding to multicolor harmonics
generation is investigated. The obtained analytical results on the basis of
generalized rotating wave approximation are in a good agreement with
numerical calculations. The results applied to the hydrogen atom and
homonuclear diatomic molecular ion show that one can achieve efficient
generation of moderately high multicolor harmonics via multiphoton resonant
excitation by appropriate laser pulses.
\end{abstract}

\maketitle



\section{ Introduction}

Harmonics generation and wave-mixing are one of the basic phenomena of
nonlinear optics which have been extensively studied both theoretically and
experimentally with the advent of lasers \cite{NO}. Recent advance in laser
technologies has provided ultrahigh intensities for supershort laser pulses
that makes achievable non-perturbative regime of harmonic generation, which
significantly extends the spectral region accessible by lasers, in
particular, for short wavelengths towards VUV/XUV or even X-ray radiation 
\cite{VUV,XUV1,XUV2,XUV3,HGrev}. Such short wavelength radiation is of great
interest due to numerous significant applications, e.g. in quantum control,
spectroscopy, sensing and imaging etc..

Depending on the laser-atom interaction parameters, harmonic generation may
arise from bound-bound \cite{BB,AAM2,AAM3,AAM1,Gib1} and bound-free-bound
transitions via continuum spectrum \cite{Huillier,BFB}. Bound-bound
mechanism of harmonic generation without ionization is more efficient for
generation of moderately high harmonics \cite{AAM2,AAM3}. For this mechanism
resonant interaction is of importance. Besides pure theoretical interest as
a simple model, resonant interaction regime exhibits significant enhancement
of frequency conversion efficiencies \cite{AAM2,AAM3}. However, to access
highly excited states of atoms/molecules by optical lasers the multiphoton
excitation problem arises. Required resonantly-driven multiphoton transition
is effective for the systems with the mean dipole moments in the stationary
states, or three-level atomic systems with close enough to each other two
states and nonzero transition dipole moment between them \cite{AM}. As a
candidate, we have studied the hydrogenlike atomic and ionic systems where
the atom has a mean dipole moment in the excited stationary states, because
of accidental degeneracy for the orbital momentum \cite{AMP,AAM4}. Other
interesting examples of efficient direct multiphoton excitation are
molecules with a permanent dipole moments \cite{Br1}, evenly charged
molecular ions at large internuclear distances \cite{Gib2}, and artificial
atoms \cite{Mer15,Oliver} realized in circuit Quantum Electrodynamics (QED)
setups \cite{CQED}.

In the work \cite{AAM4} we have shown that the multiphoton resonant
excitation of a three-level atomic/molecular system is efficient by the two
bichromatic laser fields. Hence, having efficient two-color multiphoton
resonant excitation scheme it is of interest to consider multicolor harmonic
generation and wave-mixing processes by an atomic or molecular system under
the such circumstances when only a bound states are involved in the
interaction process, which is the purpose of the current paper. The presence
of the second laser provides additional flexibility for the implementation
of multiphoton resonance expanding the spectrum of possible combinations.
Moreover, two-color excitation extends accessible scattering frequencies
with sum- and difference components. In the current paper, we employ
analytical approach for high-order multiphoton resonant excitation of
quantum systems which has been previously developed by us \cite{AM,AAM4}.
Expression for the time-dependent mean dipole moment describing coherent
part of scattering spectrum is obtained. The results based on this
expression are applied to hydrogen atom and evenly charged homonuclear
diatomic molecular ion. The main spectral characteristics of the considered
process are in good agreement with the results of the performed numerical
calculations. Estimations show that one can achieve enhanced generation of
moderately high harmonics/wave-mixing via multiphoton resonant excitation by
appropriate laser pulses. Our interest is also motivated by the advent of
circuit QED setups \cite{CQED} where one can realize artificial atoms of
desired configuration. Thus, the obtained results may also be is of interest
for artificial atoms, and the presented results can be scaled to other
systems and diverse domains of the electromagnetic spectrum.

The paper is organized as follows. In section II, we present the analytical
model and derive the coherent contribution to the multicolor harmonic
spectra. In section III, we present some results of numerical calculations
of the considered issue without a multiphoton resonant approximation and
compare the obtained spectra with the analytical results. Here, we consider
concrete systems, such as hydrogenlike atom and evenly charged molecular
ion. Finally, conclusions are given in section IV.

\section{BASIC MODEL AND ANALYTICAL ANSATZ}

We consider a three-level quantum system interacting with the two laser
fields of frequencies $\omega _{1}$ and $\omega _{2}$ as shown in Fig.(1a).
It is assumed that the system is in a $V$ configuration in which a pair of
upper levels $\left\vert 2\right\rangle $ and $\left\vert 3\right\rangle $
with permanent dipole moments are coupled to a lower level $\left\vert
1\right\rangle $. Another possible three-level scheme is $\Gamma $
configuration shown in Fig.(1b). In this case the lower level $\left\vert
1\right\rangle $ is coupled to an upper level $\left\vert 2\right\rangle $
which has a strong dipole coupling to an adjacent level $\left\vert
3\right\rangle $. If the separation of the energy levels of the excited
states is smaller than laser-atom interaction energy then by a unitary
transformation \cite{AM} the problem can be reduced to the $V$ configuration
Fig.(1a). As an illustrative example may serve hydrogenlike atom considered
in parabolic \cite{LL3} and more conventional spherical coordinates. In
parabolic coordinates, the atom has a mean dipole moment in the excited
states, while in the second case because of the random degeneracy of the
orbital moment there is a dipole coupling between the degenerate states, but
the mean dipole moment is zero for the stationary states. The inverse with
respect to the $V$ configuration is the polar $\Lambda $ configuration,
which can be realized for artificial atoms \cite{Mer15}. Hence, as a general
model we will consider the scheme of the $V$ configuration.

\begin{figure}[tbp]
\includegraphics[width=.49\textwidth]{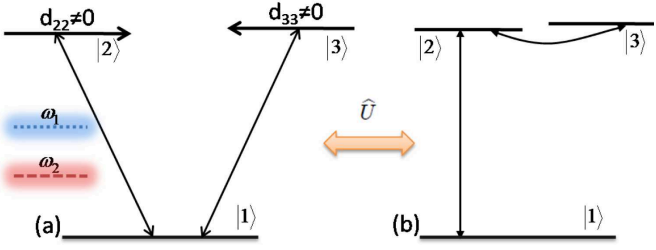}
\caption{Three-level atomic structures for (a) $V$ type with mean dipole
moments in the excited states and (b) $\Gamma $ configuration with the
coupling transition between the excited states. The considered
configurations are unitary equivalent to each other.}
\label{figg1}
\end{figure}

The Hamiltonian for the system within semiclassical dipole approximation is
given in form 
\begin{equation*}
\widehat{H}=\varepsilon _{1}|1\rangle \langle 1|+\left( \varepsilon
_{2}+V_{22}\right) |2\rangle \langle 2|+\left( \varepsilon
_{3}+V_{33}\right) |3\rangle \langle 3|
\end{equation*}%
\begin{equation}
+\left( V_{12}|1\rangle \langle 2|+V_{13}|1\rangle \langle 3|+\mathrm{h.c.}%
\right) ,  \label{Ham}
\end{equation}%
where, $\varepsilon _{1}$, $\varepsilon _{2}$ and $\varepsilon _{3}$ are the
energies of the stationary states $\left\vert 1\right\rangle $, $\left\vert
2\right\rangle $ and $\left\vert 3\right\rangle $ respectively, and%
\begin{equation}
V_{\eta \nu }=-d_{\eta \nu }\left( E_{1}\cos \left( \omega _{1}t+\varphi
\right) +E_{2}\cos \left( \omega _{2}t\right) \right) ,  \label{Interact}
\end{equation}%
is the interaction part of the Hamiltonian with a real matrix element of the
electric dipole moment projection $d_{\eta \nu }$ $=\mathbf{d}_{\eta \nu }%
\mathbf{\cdot \hat{e}}$, and $E_{1,2}$ are slowly varying amplitudes of
linearly polarized laser fields, with unit polarization vector $\mathbf{\hat{%
e}}$ and constant relative phase $\varphi $. The diagonal terms in (\ref%
{Interact}) describe the interaction due to the mean dipole moments and are
crucial for effective multiphoton coupling.

We consider Schr\"{o}dinger equation 
\begin{equation}
i\frac{\partial \left\vert \Psi (t)\right\rangle }{\partial t}=\hat{H}%
\left\vert \Psi (t)\right\rangle ,  \label{Schrod_eq}
\end{equation}%
with Hamiltonian (\ref{Ham}) at the resonance $\left\vert \delta
_{2,3}\right\vert \ll \omega _{1,2}$ for efficient multiphoton coupling. The
resonance detunings are given by relations 
\begin{equation}
\delta _{2,3}=\varepsilon _{1}-\varepsilon _{2,3}+n_{1}\omega
_{1}+n_{2}\omega _{2},  \label{detuning}
\end{equation}%
for $\left( n_{1},n_{2}\right) $ pair of photon numbers. Here and below,
unless otherwise stated, atomic units ($\hbar =e=m_{e}=1$) are employed.

Our method of solving the Schr\"{o}dinger equation with Hamiltonian (\ref%
{Ham}) has been described in detail in \cite{AAM4}, and will be excluded
here. The time-dependent wave function can be expressed as 
\begin{equation*}
|\Psi \left( t\right) \rangle =\left( \overline{a}_{1}(t)+\alpha
_{1}(t)\right) e^{-i\varepsilon _{1}t}|1\rangle
\end{equation*}%
\begin{equation*}
+\left( \overline{a}_{2}(t)+\alpha _{2}(t)\right) e^{-i\left( \varepsilon
_{2}t+\int_{0}^{t}V_{22}dt\right) }|2\rangle
\end{equation*}%
\begin{equation}
+\left( \overline{a}_{3}(t)+\alpha _{3}(t)\right) e^{-i\left( \varepsilon
_{3}t+\int_{0}^{t}V_{33}dt\right) }|3\rangle ,  \label{wf}
\end{equation}%
where $\overline{a}_{i}(t)$ are the time-averaged probability amplitudes and 
$\alpha _{i}(t)$ are rapidly changing functions on the scale of waves'
periods. Depending on the ratio of frequencies $\omega _{1}/\omega _{2}$ the
resonant condition (\ref{detuning}) can hold for a single pair of photon
numbers -normal resonance or can also be satisfied by diverse pairs of
photons numbers (in principle infinity) -degenerate resonance. Let us first
consider the case of normal resonance. Hence, if resonant condition holds
for a pair ($n,m$) then assuming the smooth turn-on of the pump waves, the
relation between rapidly oscillating and slow oscillating parts of the
probability amplitudes can be written%
\begin{equation*}
\alpha _{1}(t)=\bar{a}_{2}\sum_{\substack{ s_{1},s_{2}=-\infty ,  \\ \left(
s_{1},s_{2}\right) \neq \left( 0,0\right) }}^{\infty }\zeta
_{12}(s_{1},s_{2})e^{i\left( s_{1}\omega _{1}+s_{2}\omega _{2}\right) t}
\end{equation*}%
\begin{equation}
+\bar{a}_{3}\sum_{\substack{ s_{1},s_{2}=-\infty ,  \\ \left(
s_{1},s_{2}\right) \neq \left( 0,0\right) }}^{\infty }\zeta
_{13}(s_{1},s_{2})e^{i\left( s_{1}\omega _{1}+s_{2}\omega _{2}\right) t},
\label{rpd1}
\end{equation}%
\begin{equation}
\alpha _{2}(t)=-\bar{a}_{1}\sum_{\substack{ s_{1},s_{2}=-\infty ,  \\ \left(
s_{1},s_{2}\right) \neq \left( 0,0\right) }}^{\infty }\zeta _{12}^{\ast
}(s_{1},s_{2})e^{-i\left( s_{1}\omega _{1}+s_{2}\omega _{2}\right) t},
\label{rpd2}
\end{equation}%
\begin{equation}
\alpha _{3}(t)=-\bar{a}_{1}\sum_{\substack{ s_{1},s_{2}=-\infty ,  \\ \left(
s_{1},s_{2}\right) \neq \left( 0,0\right) }}^{\infty }\zeta _{13}^{\ast
}(s_{1},s_{2})e^{-i\left( s_{1}\omega _{1}+s_{2}\omega _{2}\right) t},
\label{rpd3}
\end{equation}%
where 
\begin{equation*}
\zeta _{12}(s_{1},s_{2})=\frac{d_{12}}{d_{22}}\frac{\left( s_{1}+n\right)
\omega _{1}+\left( s_{2}+m\right) \omega _{2}}{s_{1}\omega _{1}+s_{2}\omega
_{2}}
\end{equation*}%
\begin{equation}
\times J_{s_{1}+n}\left( \frac{d_{22}E_{1}}{\omega _{1}}\right)
J_{s_{2}+m}\left( \frac{d_{22}E_{2}}{\omega _{2}}\right) e^{i\left(
s_{1}+n\right) \varphi },  \label{z1}
\end{equation}%
and%
\begin{equation*}
\zeta _{13}(s_{1},s_{2})=\frac{d_{13}}{d_{33}}\frac{\left( s_{1}+n\right)
\omega _{1}+\left( s_{2}+m\right) \omega _{2}}{s_{1}\omega _{1}+s_{2}\omega
_{2}}
\end{equation*}%
\begin{equation}
\times J_{s_{1}+n}\left( \frac{d_{33}E_{1}}{\omega _{1}}\right)
J_{s_{2}+m}\left( \frac{d_{33}E_{2}}{\omega _{2}}\right) e^{i\left(
s_{1}+n\right) \varphi }.  \label{z2}
\end{equation}%
In deriving these equations we have applied well-known Jacobi--Anger
expansion via Bessel functions:%
\begin{equation}
e^{iZ\sin \alpha }=\sum\limits_{s=-\infty }^{\infty }J_{s}\left( Z\right)
e^{is\alpha }.  \label{bess}
\end{equation}

Now let us proceed to the case of degenerate resonance. Particularly if $%
\omega _{1}/\omega _{2}=k$, where $k$ is an integer number, then there are
many channels of resonance transitions and one should take into account all
possible transitions and the relation between rapidly oscillating and slow
oscillating parts of the probability amplitudes can be written

\begin{widetext}
\begin{equation*}
\alpha _{1}(t)=\sum_{\substack{ s_{1},s_{2}=-\infty , \\ s_{2}\neq 0}}%
^{\infty }\left[ \bar{a}_{2}\frac{d_{12}}{d_{22}}J_{s_{1}}\left( \frac{%
d_{22}E_{1}}{k\omega _{2}}\right) J_{s_{2}+n-ks_{1}}\left( \frac{d_{22}E_{2}%
}{\omega _{2}}\right) \right. 
\end{equation*}%
\begin{equation}
\left. +\bar{a}_{3}\frac{d_{13}}{d_{33}}J_{s_{1}}\left( \frac{d_{33}E_{1}}{%
k\omega _{2}}\right) J_{s_{2}+n-ks_{1}}\left( \frac{d_{33}E_{2}}{\omega _{2}}%
\right) \right] \frac{s_{2}+n}{s_{2}}e^{is_{1}\varphi }e^{is_{2}\omega
_{2}t},  \label{srpd1}
\end{equation}%
\begin{equation}
\alpha _{2}(t)=-\bar{a}_{1}\frac{d_{12}}{d_{22}}\sum_{\substack{ %
s_{1},s_{2}=-\infty , \\ s_{2}\neq 0}}^{\infty }J_{s_{1}}\left( \frac{%
d_{22}E_{1}}{k\omega _{2}}\right) J_{s_{2}+n-ks_{1}}\left( \frac{d_{22}E_{2}%
}{\omega _{2}}\right) \frac{s_{2}+n}{s_{2}}e^{-is_{1}\varphi
}e^{-is_{2}\omega _{2}t},  \label{srpd2}
\end{equation}%
\begin{equation}
\alpha _{3}(t)=-\bar{a}_{1}\frac{d_{13}}{d_{33}}\sum_{\substack{ %
s_{1},s_{2}=-\infty , \\ s_{2}\neq 0}}^{\infty }J_{s_{1}}\left( \frac{%
d_{33}E_{1}}{k\omega _{2}}\right) J_{s_{2}+n-ks_{1}}\left( \frac{d_{33}E_{2}%
}{\omega _{2}}\right) \frac{s_{2}+n}{s_{2}}e^{-is_{1}\varphi
}e^{-is_{2}\omega _{2}t},  \label{srpd3}
\end{equation}%
where $n$ is given by resonance condition%
\begin{equation}
\varepsilon _{1}-\varepsilon _{2,3}+n\omega _{2}\simeq 0.
\end{equation}%
In the Schr\"{o}dinger picture the coherent part of the dipole spectrum is
expressed as follows \cite{EF}:%
\begin{equation}
S_{c}(\omega )=\left\vert \int_{-\infty }^{\infty }dte^{-i\omega
t}\left\langle d(t)\right\rangle \right\vert ^{2},  \label{Schrod_pic}
\end{equation}%
where%
\begin{equation}
\left\langle d(t)\right\rangle =\left\langle \Psi (t)\right\vert \mathbf{%
\hat{e}\cdot \hat{d}(0)}\left\vert \Psi (t)\right\rangle ,  \label{dev1}
\end{equation}%
is the time-dependent expectation value of dipole operator. With the help of
wave function (\ref{wf}) the expectation value of the dipole operator (\ref%
{dev1}) can be written as

\begin{equation*}
\left\langle d(t)\right\rangle =\frac{d_{22}}{2}\left\vert \overline{a}%
_{2}(t)+\alpha _{2}(t)\right\vert ^{2}+\frac{d_{33}}{2}\left\vert \overline{a%
}_{3}(t)+\alpha _{3}(t)\right\vert ^{2}
\end{equation*}%
\begin{equation*}
+d_{12}\left( \overline{a}_{1}^{\ast }(t)+\alpha _{1}^{\ast }(t)\right)
\left( \overline{a}_{2}(t)+\alpha _{2}(t)\right)
\sum\limits_{s_{1},s_{2}=-\infty }^{\infty }J_{s_{1}}\left( \frac{d_{22}E_{1}%
}{\omega _{1}}\right) J_{s_{2}}\left( \frac{d_{22}E_{2}}{\omega _{2}}\right)
e^{i\left( s_{1}\omega _{1}+s_{2}\omega _{2}\right) t}e^{i\left( \varepsilon
_{1}-\varepsilon _{2}\right) t+is_{1}\varphi }
\end{equation*}%
\begin{equation}
+d_{13}\left( \overline{a}_{1}^{\ast }(t)+\alpha _{1}^{\ast }(t)\right)
\left( \overline{a}_{3}(t)+\alpha _{3}(t)\right)
\sum\limits_{s_{1},s_{2}=-\infty }^{\infty }J_{s_{1}}\left( \frac{d_{33}E_{1}%
}{\omega _{1}}\right) J_{s_{2}}\left( \frac{d_{33}E_{2}}{\omega _{2}}\right)
e^{i\left( s_{1}\omega _{1}+s_{2}\omega _{2}\right) t}e^{i\left( \varepsilon
_{1}-\varepsilon _{3}\right) t+is_{1}\varphi }+\mathrm{c.c.}.  \label{dev2}
\end{equation}%
\end{widetext}
Combining the solution for slow oscillating parts of the probability
amplitudes with (\ref{z1}), (\ref{z2}) and (\ref{dev2}) one can calculate
analytically the expectation value of the dipole operator for an arbitrary
initial atomic state. The Fourier transform of $\left\langle
d(t)\right\rangle $ gives the coherent part of the dipole spectrum.

The solution for slow oscillating parts of the probability amplitudes
analytically is very complicated and in order to reveal the physics of
multiphoton resonant excitation process let us consider systems with
inversion symmetry: $d_{12}=-d_{13}\equiv d_{tr}$ and $d_{22}=-d_{33}\equiv
d $. As is seen from (\ref{dev2}), for effective harmonic generation in
certain resonance conditions one should provide considerable population
transfer between the atomic states $\left\vert 1\right\rangle $ and upper
levels $\left\vert 2\right\rangle $ and $\left\vert 3\right\rangle $.
Dynamic Stark shifts can take the states off\ resonance, so appropriate
detunings for compensation are chosen.

Let us first consider the case of normal resonance. For the system initially
situated in the ground state, the solution for slow oscillating parts of the
probability amplitudes is \cite{AAM4}: 
\begin{equation}
\overline{a}_{1}(t)=e^{-i2\Delta t}\sin (\Omega _{R}t),  \label{avg1}
\end{equation}%
\begin{equation}
\overline{a}_{2}(t)=\frac{(-1)^{n+m}}{i\sqrt{2}}e^{-i2\Delta t}\sin (\Omega
_{R}t),  \label{avg2}
\end{equation}%
\begin{equation}
\overline{a}_{3}(t)=\frac{1}{i\sqrt{2}}e^{-i2\Delta t}\sin (\Omega _{R}t),
\label{avg3}
\end{equation}%
where $\Delta $ describes dynamic Stark shifts%
\begin{equation*}
\Delta \equiv \left( \frac{d_{tr}}{d}\right) ^{2}\sum_{\substack{ %
s_{1},s_{2}=-\infty ,  \\ \left( s_{1},s_{2}\right) \neq \left( n,m\right) }}%
^{\infty }\frac{\left( s_{1}\omega _{1}+s_{2}\omega _{2}\right) ^{2}}{%
(s_{1}-n)\omega _{1}+(s_{2}-m)\omega _{2}}
\end{equation*}%
\begin{equation}
\times J_{s_{1}}^{2}\left( \frac{dE_{1}}{\omega _{1}}\right)
J_{s_{2}}^{2}\left( \frac{dE_{2}}{\omega _{2}}\right) .  \label{Stark}
\end{equation}%
and $\Omega _{R}$ expresses the frequency of Rabi oscillations 
\begin{equation}
\Omega _{R}\equiv \left\vert 2\sqrt{2}\frac{d_{tr}}{d}J_{n}\left( \frac{%
dE_{1}}{\omega _{1}}\right) J_{m}\left( \frac{dE_{2}}{\omega _{2}}\right)
\right\vert .  \label{Rabi1}
\end{equation}%
\ Replacing the probability amplitudes in (\ref{dev2}) by the corresponding
expressions (\ref{avg1})-(\ref{avg3}) and taking into account the relations (%
\ref{rpd1})-(\ref{rpd3}) one can derive the final analytical expression for $%
\left\langle d(t)\right\rangle $. As is seen from Eq. (\ref{Stark}), the
dynamic Stark shift is proportional to the ratio $d_{tr}^{2}/d^{2}$, while
the multiphoton coupling is proportional to $d_{tr}/d$. Since large dynamic
Stark shifts are detrimental for maintenance of considerable population
transfer, here we consider systems with $\left\vert d_{tr}/d\right\vert \ll
1 $. Taking into account the smallness of the parameter $\left\vert
d_{tr}/d\right\vert $, from (\ref{dev2}) at the first approximation we
derive the following compact analytic formula:%
\begin{equation}
\left\langle d(t)\right\rangle =\sum_{\substack{ s_{1},s_{2}=-\infty ,  \\ %
\left( s_{1},s_{2}\right) \neq \left( 0,0\right) }}^{\infty
}D_{s_{1}s_{2}}\sin ((s_{1}\omega _{1}+s_{2}\omega _{2})t+s_{1}\varphi ),
\label{main}
\end{equation}%
where%
\begin{equation*}
D_{s_{1}s_{2}}=\frac{d_{tr}}{\sqrt{2}}\sin (\Omega _{R}t)\frac{n\omega
_{1}+m\omega _{2}}{s_{1}\omega _{1}+s_{2}\omega _{2}}(1-(-1)^{s_{1}+s_{2}})
\end{equation*}%
\begin{equation}
\times J_{s_{1}+n}\left( \frac{dE_{1}}{\omega _{1}}\right) J_{s_{2}+m}\left( 
\frac{dE_{2}}{\omega _{2}}\right) .  \label{amp}
\end{equation}

As we can see from (\ref{main}), (\ref{amp}), the spectrum consists of
doublets $s_{1}\omega _{1}+s_{2}\omega _{2}\pm \Omega _{R}$. At that only
harmonics with odd sum of harmonic numbers $s_{1}+s_{2}$ are existed, as it
was expected because of inversion symmetry of the considered problem.

Now let us consider the degenerate resonance. Particularly if $\omega
_{1}/\omega _{2}=k$, where $k$ is odd, the dipole expectation value can be
obtained from Eqs. (\ref{srpd1})-(\ref{srpd3}), (\ref{dev2}) and is given
as: 
\begin{equation*}
\left\langle d(t)\right\rangle =\frac{d_{tr}n}{\sqrt{2}}\sum_{\substack{ %
s_{1},s_{2}=-\infty ,  \\ s_{2}\neq 0}}^{\infty }J_{s_{2}+n-ks_{1}}\left( 
\frac{dE_{2}}{\omega _{2}}\right) J_{s_{1}}\left( \frac{dE_{1}}{k\omega _{2}}%
\right)
\end{equation*}%
\begin{equation}
\times \frac{1-(-1)^{s_{2}}}{s_{2}}\sin (\Omega _{R}^{\prime }t)\sin
(s_{2}\omega _{2}t+s_{1}\varphi ),  \label{main_D}
\end{equation}%
where Rabi frequenc is given as:%
\begin{equation}
\Omega _{R}^{\prime }\equiv \left\vert 2\sqrt{2}\frac{d_{tr}}{d}n\omega
_{2}\sum\limits_{s}J_{n-ks}\left( \frac{dE_{2}}{\omega _{2}}\right)
J_{s}\left( \frac{dE_{1}}{k\omega _{2}}\right) \right\vert .  \label{Rabi_D}
\end{equation}

\ The spectrum is noticeably different for even $k$. In particular, for
bichromatic field with frequencies $\omega $ and $2\omega $ dipole spectrum
contains also even harmonics of the main frequency, tunable low-frequency
(much smaller than $\omega $) part and six hyper-Raman lines per harmonic.
Dipole moment expectation value is expressed by very bulky formula and will
not be given here.

\section{NUMERICAL RESULTS AND\ DISCUSSION}

In this section we present numerical calculations for hydrogen atom and
homonuclear diatomic molecular ion $N_{2}^{4+}$ \cite{Gib2} with the
specific parameters of available laboratory lasers. The numerical results
will be compared with exact results of the dipole spectrum to estimate
accuracy and applicability of generalized rotating wave approximation. The
time-dependent Schr\"{o}dinger equation for three-level model with
Hamiltonian (\ref{Ham}) are considered. The solution for the probability
amplitudes has been obtained using Runge-Kutta algorithm and scattering
spectrum is estimated by applying the fast Fourier transform method \cite{NR}%
. For smooth turn-on of the laser fields, we consider envelopes with
hyperbolic tangent $\tanh (t/\tau )$ temporal shape, where $\tau $
characterizes the turn-on time and chosen to be $40\pi /\omega _{2}$. It is
assumed that quantum systems are initially in the ground state $\left\vert
1\right\rangle $. For both systems the main transitions fall in the vacuum
ultraviolet range. Thus, $\varepsilon _{1}-\varepsilon _{2,3}\simeq 0.375$ $%
\mathrm{a.u}$. and $\simeq 0.6838$ $\mathrm{a.u}$. for hydrogen atom and ion 
$N_{2}^{4+}$, respectively. This is a spectral domain where strong coherent
radiation is difficult to generate and two or higher photon multicolor
resonant excitation is of interest. 
\begin{figure}[tbp]
\includegraphics[width=.49\textwidth]{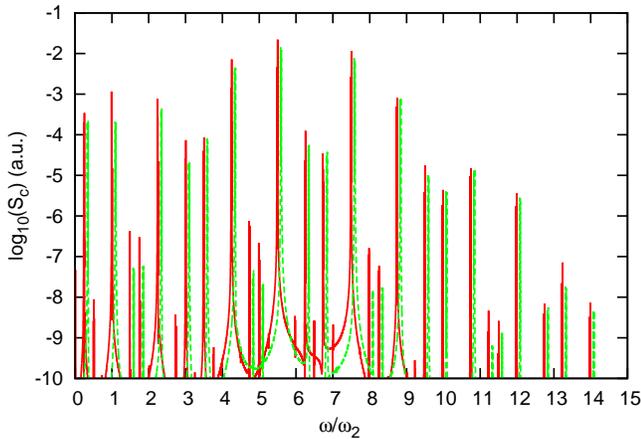}
\caption{(Color online) The logarithm of the coherent part of the spectrum $%
S_{C}(\protect\omega )$ at four-photon two-color resonance ($n=2,m=2$) of
hydrogen atom with $E_{1}=E_{2}=0.02$ $\mathrm{a.u}$\textrm{, }$\protect%
\omega _{1}=0.13$ $\mathrm{a.u}$, and $\protect\omega _{2}=0.0579$ $\mathrm{%
a.u.}$ The solid (red) line corresponds to numerical calculations; the
dashed (green) line corresponds to the approximate solution (for better
visibility the latter has been slightly shifted to the right). }
\end{figure}
\begin{figure}[tbp]
\includegraphics[width=.49\textwidth]{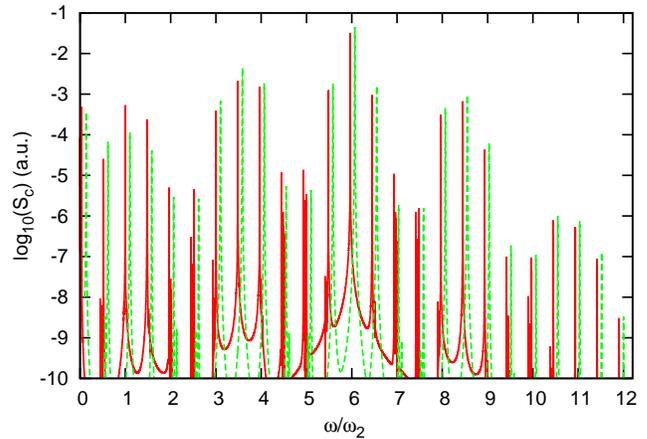}
\caption{(Color online) The logarithm of the coherent part of the spectrum $%
S_{C}(\protect\omega )$ at five-photon two-color resonance ($n=2,m=3$) of
hydrogen atom with $E_{1}=E_{2}=0.02$ $\mathrm{a.u}$\textrm{, }$\protect%
\omega _{1}=0.0935$ $\mathrm{a.u}$, and $\protect\omega _{2}=0.063$ $\mathrm{%
a.u.}$ The solid (red) line corresponds to numerical calculations; the
dashed (green) line corresponds to the approximate solution.}
\label{amp23}
\end{figure}

Figure 2 displays the multicolor harmonic and wave-mixing emission rate
(coherent part) as a function of the ratio $\omega /\omega _{2}$ (we assume $%
\omega _{2}<\omega _{1}$ and $\varphi =0$) at the four-photon two-color
resonant excitation of hydrogen atom with XeF excimer (351 nm, $n=2$) and
Ti:sapphire (780 nm, $m=2$) laser systems with $E_{1}=E_{2}=0.02$ $\mathrm{%
a.u}$.. For the hydrogen atom $d_{tr}=0.5267$ $\mathrm{a.u}$. and $d=3.0$ $%
\mathrm{a.u}$. Here and below, for the chosen parameters the dynamic Stark
shift is compensated. The latter provides almost complete population
transfer. The solid (red) line corresponds to numerical calculations, while
the dashed (green) line corresponds to the approximate expression (\ref{main}%
). Note that the numerical and analytical calculations coincide with high
accuracy, so for visual convenience to distinguish these curves the spectrum
corresponding to analytical calculations (\ref{main}) has been slightly
shifted to the right for Figs. 2-5.

Figure 3 displays the multicolor harmonic and wave-mixing emission rate at
the five-photon two-color resonant excitation of hydrogen atom with Ar+ (488
nm, $n=2$) and Ti:sapphire (724 nm, $m=3$) laser systems with $%
E_{1}=E_{2}=0.02$ $\mathrm{a.u}$. 
\begin{figure}[tbp]
\includegraphics[width=.49\textwidth]{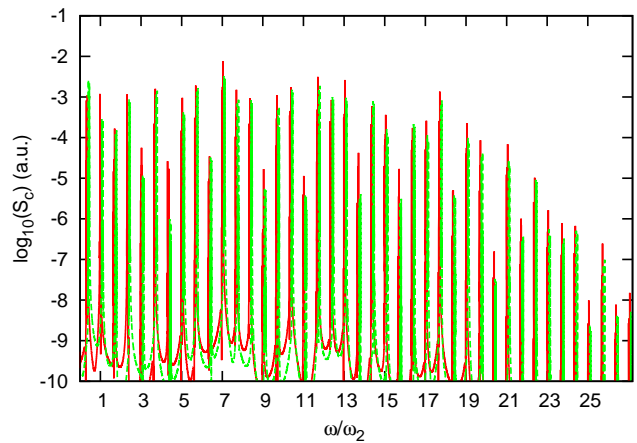}
\caption{(Color online) The logarithm of the coherent part of the spectrum $%
S_{C}(\protect\omega )$ at seven-photon two-color resonance ($n=4,m=3$) of N$%
_{2}^{4+}$ molecular ion with $E_{1}=E_{2}=0.07$ $\mathrm{a.u}$\textrm{, }$%
\protect\omega _{1}=0.13$ $\mathrm{a.u}$, and $\protect\omega _{2}=0.05562$ $%
\mathrm{a.u.}$ The solid (red) line corresponds to numerical calculations;
the dashed (green) line corresponds to the approximate solution.}
\end{figure}
\begin{figure}[tbp]
\includegraphics[width=.49\textwidth]{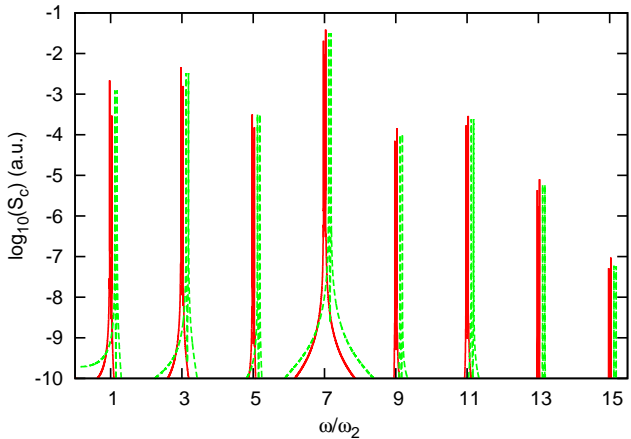}
\caption{(Color online) The radiation spectrum at two-color resonant
excitation of hydrogen atom with laser fields $E_{1}=E_{2}=0.02$ $\mathrm{a.u%
}$\textrm{, }$\protect\omega _{1}=3\protect\omega _{2}$, and $\protect\omega %
_{2}=0.05371$ $\mathrm{a.u..}$}
\end{figure}

In Fig. 4 we plot harmonic/wave-mixing emission rate at the seven-photon
two-color resonant excitation of N$_{2}^{4+}$ molecular ion with XeF excimer
(351 nm, $n=4$) and Ti:sapphire (820 nm, $m=3$) laser systems with $%
E_{1}=E_{2}=0.07$ $\mathrm{a.u}$. For this system we take $d_{tr}=0.3536$ $%
\mathrm{a.u}$. and $d=3.0$ $\mathrm{a.u}$..

For a degenerate case of resonance in Fig. 5 we plot harmonic/wave-mixing
emission rate at $\omega _{1}=3\omega _{2}$ with $\omega _{2}=0.05371$ $%
\mathrm{a.u}$. (849nm) at waves' electric fields $E_{1}=E_{2}=0.02$ $\mathrm{%
a.u}$.. For the case $\omega _{1}=2\omega _{2}$ the radiation spectrum is
similar to non-degenerate case. However, there are a few differences.
Scattering spectrum is richer in satellites of multicolor harmonics, in
addition there is low-frequency radiation on Rabi frequencies. In Fig. 6 we
plot low-frequency part of radiation spectrum. Here, the presented triplet
which can be tuned by laser parameters lies in THz/IR region. 
\begin{figure}[tbp]
\includegraphics[width=.49\textwidth]{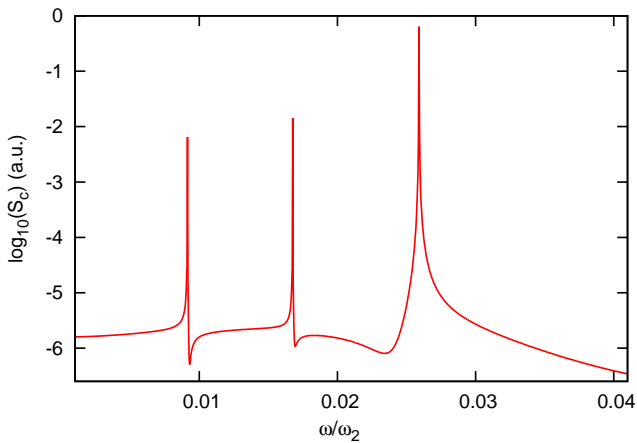}
\caption{(Color online) Low-frequency part of radiation spectrum at
two-color resonant excitation of hydrogen atom with laser fields $%
E_{1}=E_{2}=0.025$ $\mathrm{a.u}$\textrm{, }$\protect\omega _{1}=2\protect%
\omega _{2}$, and $\protect\omega _{2}=0.06267$ $\mathrm{a.u..}$ }
\end{figure}

As is seen from these figures in the coherent spectrum there are as
harmonics of the individual waves as well as frequencies with sum/difference
components and its harmonics, in accordance with the analytical ansatz (\ref%
{main_D}). From Eq. (\ref{amp}) it is clear that for effective harmonic
generation one should provide large dipole interaction energy $%
dE_{1,2}\gtrsim \omega _{1,2}$, since the Bessel function exponentially
decreases with increasing index at the given argument. The Bessel function $%
J_{s}\left( Z\right) $ at large argument values reaches its maximum at $%
s\sim Z$. Thus, the cutoff frequency depends linearly on the amplitudes of
the laser fields.

Let us make some estimations for reasonable interaction parameters. The
average number of photons at the frequency $\omega $ emitted at each lasers
shots of duration $\tau $, Rayleigh length $L_{R}$ on the atomic/molecular
ensemble of density $N_{a}$ can be estimated as \cite{AAM2}: 
\begin{equation*}
N_{\omega }\simeq \frac{\left( 2\pi \right) ^{2}}{\hbar }S_{c}(\omega
)N_{a}^{2}L_{R}^{3}\tau
\end{equation*}%
The incident pulse duration is assumed to be $50$\ $\mathrm{ps}$, the
Rayleigh length is taken to be $L_{R}=1\ \mathrm{mm}$, and for emitters
density we assume $N_{a}\simeq 5\times 10^{17}$\textrm{cm}$^{-3}$. For the
setup of Figs. 3 and 4 with the chosen parameters, the average number of
radiated photons at frequencies up to $\omega \simeq 20\omega _{2}$ per shot
is $N_{\omega }\sim 10^{12}$, which is two orders of magnitude larger than
what one expects to achieve with tunneling harmonics generation \cite%
{Huillier}.

\section{SUMMARY}

We have presented a theoretical treatment of the multicolor harmonics
generation and wave-mixing in a three-level atomic-molecular system under
two-color multiphoton resonant excitation. The coherent part of the dipole
spectrum was investigated. With the help of an approximate analytical
expression for the dynamic wave function of a three-level atom driven by
intense laser fields, we obtained an analytical expression for the
time-dependent expectation value of the dipole operator. Then the results
obtained were applied to the hydrogen atom and homonuclear diatomic
molecular ion. The spectrum shows harmonics of the individual waves as well
as frequencies with sum/difference components and its harmonics. These peaks
have quite large amplitudes. The latter is the result of multiphoton
resonant interaction of the system with the driving bichromatic laser
radiation due to the mean dipole moment in the stationary states. The cutoff
frequency depends linearly on the amplitudes of laser fields. The presence
of the second laser field can make easier the implementation of efficient
population transfer and harmonic generation as well as allows generation of
new frequencies. Analytical calculations in the generalized rotating wave
approximation \cite{AM,AAM4} allow an explanation of the obtained spectrum.
The numerical simulations are in good agreement with the analytical results.
The considered scheme may serve as a promising method for efficient
production of multicolor high harmonics. It should be noted that the
obtained results can be applied to other systems in diverse domains of the
electromagnetic spectrum.

\begin{acknowledgments}
This work was supported by the RA MES State Committee of Science, in the
frames of the research project No. 15T-1C013.
\end{acknowledgments}

\end{document}